# Proton spin dynamics in polymer melts: new perspectives for experimental investigations of polymer dynamics.

**N. Fatkullin[a], S. Stapf[b], M. Hofmann[c], R. Meier[c], E.A. Rössler[c]**

[a] Institute of Physics, Kazan Federal University, Kazan, 42008 Tatarstan, Russia
[b] Dept. Technical Physics II/Polymer Physics, TU Ilmenau, PO Box 100 565, D-98684 Ilmenau, Germany
[c] Universität Bayreuth, Experimentalphysik II, D-95440 Bayreuth, Germany

The proton spin dynamics in polymer melts is determined by intramolecular and intermolecular magnetic dipole-dipole contributions of proton spins. During many decades it was postulated that the main contribution is a result of intramolecular magnetic dipole-dipole interactions of protons belonging to the same polymer segment. This postulate is far from reality. The relative weights of intra- and intermolecular contributions are time (or frequency) dependent and sensitive to details of polymer chain dynamics. It is shown that for isotropic models of polymer dynamics, in which already at short times the segmental displacements are not correlated with the polymer chain's initial conformation, the influence of the intermolecular dipole-dipole interactions becomes stronger with increasing evolution time (i.e. decreasing frequency) than the corresponding influence of the intramolecular counterpart. On the other hand, an inverted situation is predicted by the tube-reptation model: here the influence of the intramolecular dipole-dipole interactions increases faster with time than the contribution from intermolecular interactions. This opens a new perspective for experimental investigations of polymer dynamics by proton NMR, and first results are reported.

## 1. Introduction.

Proton NMR is a powerful method for experimental investigations of structure and dynamics in different fields of condensed matter in general, and polymer physics in particular [1-13]. This favorable situation is determined at least by the two following facts. Firstly, protons are ubiquitous and are present in the majority of soft matter of interest. Secondly, the dynamics of the proton spin, as of any another spin nucleus, in an external magnetic field is simple and exactly solvable using rather elementary mathematics. In the absence of other interactions the nuclear spin performs precession around the Z axis along which the magnetic field is aligned with the Larmor frequency given by the simple relation:

$$\omega = -\gamma_H B_0, \qquad (1)$$

where $B_0$ is the experimentally controlled external magnetic field, $\gamma_H$ is the gyromagnetic ratio of the proton. Interactions of protons with each other as well as other degrees of freedom disturb the simple picture. A quantitative description of these various influences is the main subject of NMR theory in condensed matter. Additional interactions induce a shift of the proton frequency and create relaxation processes. Shifts of the resonance frequency, the most important of which is the chemical shift generated by electronic shielding, are the main subject of NMR spectroscopy [1-3] and form a basis for studying the microscopic structure of polymers based on the experimentally observed NMR spectra. The dynamics of the investigated systems is mainly reflected through characteristic features of nuclear spin relaxations. Important findings of recent years [7,11-30], affecting the understanding of the proton spin dynamics in polymer melts are the subject of discussion of this paper.

One has to distinguish relaxation parallel and perpendicular to the quantization ($Z$) axis, which is defined by the direction of the external magnetic field. The longitudinal relaxation, i.e. the spin relaxation along $Z$ direction, is characterized by the spin-lattice relaxation time $T_1(\omega)$. The transverse relaxation, i.e., the spin relaxation in the $XY$ plane is given by the spin-spin relaxation time $T_2(\omega)$. For polymer systems with large molecular masses, $T_1(\omega)$ possesses a non-trivial frequency dispersion covering an extremely broad frequency range which nowadays can be measured most easily by field cycling techniques in a frequency range of 100 Hz – 40 MHz when earth field compensation is included [24,25,30,32]. The relaxation time $T_2(\omega)$ has weaker frequency dependence and is usually investigated at fixed resonance frequency. As a rule, $T_1(\omega) \geq T_2(\omega)$.

The main interaction controlling spin relaxation in polymer systems is the magnetic dipole-dipole interaction between different protons in the system, the Hamiltonian of which is as follows[1]:

$$\hat{H}_{dd} = \sum_{i<j} \frac{\mu_0}{4\pi} \frac{\gamma_H^2 \hbar^2}{r_{ij}^3} \left\{ \hat{\vec{I}}_i \cdot \hat{\vec{I}}_j - 3\left(\hat{\vec{I}}_i \cdot \vec{e}_{ij}\right)\left(\hat{\vec{I}}_j \cdot \vec{e}_{ij}\right) \right\}, \quad (2)$$

where $r_{ij}$ is the internuclear distance, $\vec{e}_{ij} = \vec{r}_{ij}/r_{ij}$, $\hat{\vec{I}}_i$ is the spin vector operator of spin number $i$, and $\mu_0$ is the magnetic field constant. Summation in the expression (2) is to be taken over all pairs of spins. To be more specific, summation is performed over both spins belonging to the same macromolecule, which constitutes the intramolecular contribution, and spins from different macromolecules representing the intermolecular contribution.

The dynamics of spin relaxation is determined by the second order corrections arising from the Hamiltonian (2). More precisely, the initial part of the spin relaxation at times $t \leq T_1(\omega), T_2(\omega)$ is fully determined by the following time dependent dipolar correlation functions:

$$A_p(t) = \frac{1}{N_s} \sum_{k \neq m} \left\langle \frac{Y_{2p}(\vec{e}_{km}(t))}{r_{km}^3(t)} \frac{Y_{2p}^*(\vec{e}_{km}(0))}{r_{km}^3(0)} \right\rangle, \quad (3)$$

where $Y_{lp}(\vec{e}_{ij})$ is component $p$ of the spherical function of rank $l$, $N_s$ is the number of spins in the system, and $\vec{e}_{km}(t) = \frac{\vec{r}_{km}(t)}{r_{km}(t)}$; here $\vec{r}_{km}(t)$ are the internuclear vectors. In isotropic systems like polymer melts, the correlation function $A_p(t)$ does not depend on $p$, and all components are equal, i.e. $A_0(t) = A_1(t) = A_2(t)$. To illustrate the connection of the correlation function $A_p(t)$ with experimentally measurable quantities let us briefly discuss its connection with the free induction decay (FID), the spin-lattice relaxation rate and the normalized double quantum proton kinetic build up curve.

The free induction decay (FID) is conceptually the simplest NMR phenomenon. It is generated after applying to the equilibrium spin system in a magnetic field a radiofrequency pulse which rotates spins about an angle $\pi/2$ around an axis perpendicular to $Z$. In the framework of the Anderson-Weiss approximation the FID can be expressed through $A_0(t)$ [1,29], in the following way:

$$g(t) = \exp\left\{ -\left(\frac{\mu_0}{4\pi}\right)^2 \frac{9\pi}{5} \gamma_H^4 \hbar^2 \int_0^t d\tau (t-\tau) A_0(\tau) \right\}. \quad (4)$$

Recently using a modified Anderson-Weiss approximation, taking into account the spin diffusion process, i.e., transfer of longitudinal magnetization from one spin to another, it was shown that approximation (4) gives an error smaller than 10%, for times $t \leq 2T_2$ [22].

The the spin-lattice relaxation rate is a linear combination of Fourier transforms of $A_p(t)$ at the resonance and the double resonance frequencies [1-6,11]:

$$\frac{1}{T_1(\omega)} = \left(\frac{\mu_0}{4\pi}\right)^2 \frac{6\pi}{5} \gamma_H^4 \hbar^2 \int_0^\infty \left\{\cos(\omega t) A_1(t) + 4\cos(2\omega t) A_2(t)\right\} dt. \tag{5}$$

Note this expression is correct when the so-called short correlation time approximation, or Redfield limit, is satisfied, i.e. $T_1(\omega) \gg \min\{\omega^{-1}, \tau_1\}$, where $\tau_1$ is the terminal relaxation time, i.e., the longest relaxation time in a polymer melt.

The double quantum proton NMR (DQ NMR) can be characterized as a response of the spin system to the particular DQ pulse sequence [8,22,30,34-35] which effectively transforms the Hamiltonian of the dipole-dipole interactions in the rotating frame to the DQ Hamiltonian. The which induces only even spin transitions, among which at times $t \leq 2T_2$ the two-spin transitions dominate. Experimentally measured data employing this method can be expressed through the so-called normalized DQ kinetic (build-up) curve. The DQ proton kinetic curve is, again in the framework of the Anderson-Weiss approximation, also fully determined by the correlation function $A_0(t)$ [29]:

$$I_{nDQ}(\tau_{DQ}) = \frac{1}{2}\left(1 - \exp\left\{-\left(\frac{\mu_0}{4\pi}\right)^2 \frac{8\pi}{5} \gamma_H^4 \hbar^2 \int_0^{\tau_{DQ}} d\tau (\tau_{DQ} - \tau)\left(A_0(\tau_{DQ} + \tau) + A_0(\tau_{DQ} - \tau)\right)\right\}\right), \tag{6}$$

where $\tau_{DQ}$ is the experimentally controlled, DQ excitation time. In the next chapters spin-lattice relaxation is discussed in more detail.

## 2. Main features of the total time-dependent magnetic dipole-dipole correlation decay in polymer melts.

In polymer melts, protons are mobile. At short times $t < \tau_s$, where $\tau_s$ is the polymer segment relaxation time, relative displacements of protons are much smaller than their initial relative distances. At this situation just like in the solid state, the main contribution to the correlation function $A_p(t)$ stems from nearest intramolecular protons belonging to the same Kuhn segment. At longer times, the situation is qualitatively different, because intra- and intermolecular contributions to $A_p(t)$ decay in different ways, which depend on the detail of polymer dynamics at times $\tau_s < t < \tau_1$. Note that in entangled polymer melts, the terminal relaxation time strongly increases with polymer chain length (or molecular mass), explicitly $\tau_1 \propto \tau_s N^{3.4}$, where $N$ is the number of Kuhn segments per chain; times on the order of seconds or more are found for long chains.

The total dipolar correlation function can be represented as a sum of inter- and intramolecular components:

$$A_p(t) = A_p^{inter}(t) + A_p^{intra}(t). \tag{7}$$

We will discuss the two contributions separately.

### 2.1 Intermolecular relaxation contribution

Let us begin by discussing the intermolecular part $A_p^{inter}(t)$, which for times $t \gg \tau_s$ has a universal form valid for all dynamic models of polymer melts [16]:

$$A_p^{inter}(t) = \frac{4\pi}{9} n_s \tilde{W}(0;t), \tag{8}$$

where $n_s$ is the concentration of proton spins, and $\tilde{W}(\vec{r}',\vec{r};t) = \tilde{W}(\vec{r}'-\vec{r};t)$ is the propagator of relative displacements between two spins on different macromolecules, i.e., the density of probability for two spins separated initially by a vector $\vec{r}$ to be separated by a vector $\vec{r}'$ after an interval $t$. In fact, the intermolecular relaxation contribution for times $t \gg \tau_s$ is proportional to the probability density to recover after an interval $t$ their initial spatial separation, i.e., the vector $\vec{r}$, which is not necessarily small and can be arbitrary.

The next step of analysis of (8) is the approximation of the propagator by a Gaussian distribution [16, 21]:

$$\tilde{W}(r;t) = \frac{1}{\left((2\pi/3)\langle \tilde{r}^2(t) \rangle\right)^{3/2}} \exp\left\{-\frac{3}{2}\frac{r^2}{\langle \tilde{r}^2(t) \rangle}\right\}, \qquad (9)$$

where $\langle \tilde{r}^2(t) \rangle = \langle (\vec{r}' - \vec{r})^2 \rangle$ is the relative mean-squared displacement of two spins belonging to different macromolecules. With this, and setting $r = 0$, eq. (8) turns into the following expression [16, 21]:

$$A_p^{\text{inter}}(t) = \sqrt{\frac{2}{3\pi}} \frac{n_s}{\langle \tilde{r}^2(t) \rangle^{3/2}}. \qquad (10)$$

## 2.2 Intramolecular relaxation contribution

The dependence of the intramolecular part $A_p^{\text{intra}}(t)$ in (7) does not possess the same universal character as in (10), and depends on the details of polymer dynamics which is different for so-called isotropic models like the n-renormalized Rouse and the polymer mode-mode coupling model [36-40], and anisotropic approaches like the tube-reptation model [41-44]. Note that the first two models are isotropic as they assume that already at short times $t \gg \tau_s$ the segmental displacements are not correlated with the polymer chain's initial conformation. In contrast, the tube-reptation model, where in the time interval $\tau_e \leq t \leq \tau_1$ spatial displacements of polymer segments are confined inside the tube ($\tau_e$ is the entanglement time), is an anisotropic polymer model since conformations are strongly correlated with the polymer chain's initial conformation.

The contribution $A_p^{\text{intra}}(t)$ can itself be split into the intrasegment contribution $A_p^s(t)$ referring to spin pairs within a Kuhn segment, and the intersegment or segment-segment contribution $A_p^{ss}(t)$ implying pairs of spins in different segments of the same macromolecule:

$$A_p^{\text{intra}}(t) = A_p^s(t) + A_p^{ss}(t). \qquad (11)$$

The main contribution in (11) for all types of polymer models in the experimentally accessible time/frequency window arises from the intrasegment contribution $A_p^s(t)$. For times $t \gg \tau_s$ the correlation function has the following structure [20,21]:

$$A_0^s(t) = \frac{5}{16\pi} \tilde{S} \left\langle \left[ \left( b_k^2(t) \right) - 3\left( b_k^z(t) \right)^2 \right] \left[ \left( b_k^2(0) \right) - 3\left( b_k^z(0) \right)^2 \right] \right\rangle, \tag{12}$$

where $\tilde{S}$ is a coefficient depending on the chemical structure of the macromolecule, $b_k^2(t) = \left( b_k^x(t) \right)^2 + \left( b_k^y(t) \right)^2 + \left( b_k^z(t) \right)^2$ is the square of the Kuhn segment length with number $k$ at time $t$, $b_k^\alpha(t)$ is the component $\alpha = x, y, z$ of the vector connecting the ends of the Kuhn segment at time $t$, the bracket $\langle ... \rangle$ denotes the averaging with respect to an equilibrium distribution function and averaging over all Kuhn segments. The vector $\vec{b}_k$ can be considered as the tangent vector of the polymer chain at the position of the $k$-th Kuhn segment and the corresponding expression in the large bracket of (12) is a rank-two reorientational correlation function. The factor $\tilde{S}$ has the following structure [21]:

$$\tilde{S} = \frac{1}{\tilde{N}_s} \sum_{i \neq j} \frac{S_{ij}^2}{\langle b_k^4 \rangle^2} \left\langle \frac{b_k^2}{r_{k;ij}^3} \right\rangle^2, \tag{13}$$

where summation is performed over all spins belonging to the same Kuhn segment, $\tilde{N}_s$ is the number of spins per Kuhn segment, $r_{ij}$ is the distance between spins with numbers $i$ and $j$, $i \neq j$ $S_{ij}$ is the spin-segment coupling constant. The coupling constant between spins depends on the chemical structure of the macromolecule and can be expressed through equilibrium static correlation functions in the following way:

$$S_{ij} = \frac{1}{2} \left\langle \frac{3\left( \vec{e}_{k;ij} \cdot \vec{n}_k \right)^2 - 1}{r_{k;ij}^3} b_k^2 \right\rangle \left\langle \frac{b_k^2}{r_{k;ij}^3} \right\rangle^{-1}, \tag{14}$$

where $\vec{n}_k = \vec{b}_k / b_k$ is the unit vector parallel to the tangent vector $\vec{b}_k$. We note that $S_{ij}$ is sometimes also called residual dipolar coupling or order parameter. The parameter $\tilde{S}$ takes into account the reduction of the dipolar interaction due to fast local motions, i.e., in more general terms, fast thermal fluctuations of intra-segmental degrees of freedom at times $t \leq \tau_s$. For protons in the $CH_2$ group, for example, $S_{CH_2} \approx -\frac{1}{2}$, while for protons in $CH_3$ groups $S_{CH_3} \approx \frac{1}{4}$.

The most important factor in (12) containing information about polymer chain dynamics at times $t \gg \tau_s$ is the fourth order, or four-point time correlation function:

$$B_0(t) = \left\langle \left[ \left(b_k^2(t)\right) - 3\left(b_k^z(t)\right)^2 \right] \left[ \left(b_k^2(0)\right) - 3\left(b_k^z(0)\right)^2 \right] \right\rangle. \tag{15}$$

The relation between this four-point correlation function with the more elementary two-point correlation functions is qualitatively different between isotropic polymer models and the tube-reptation model. As already discussed, in isotropic polymer models it is postulated that different components of the vector $\vec{b}_k$ fluctuate independently from each other and the fourth order correlation function is therefore proportional to the square of the binary correlation function:

$$\left\langle \left[ \left(b_k^2(t)\right) - 3\left(b_k^z(t)\right)^2 \right] \left[ \left(b_k^2(0)\right) - 3\left(b_k^z(0)\right)^2 \right] \right\rangle^{isot} = \frac{4}{3} \left\langle \vec{b}_k(t) \vec{b}_k(0) \right\rangle^2. \tag{16}$$

Over a very wide range of time, $\tau_s \ll t \ll \tau_1$, the tangent vector correlation function $\left\langle \vec{b}_k(t) \cdot \vec{b}_k(0) \right\rangle$ is simply connected with the segmental mean squared displacement [45]:

$$\left\langle r_k^2(t) \right\rangle \left\langle \vec{b}_k(t) \cdot \vec{b}_k(0) \right\rangle = \beta \frac{b^4}{\pi}, \tag{17}$$

where $b = \sqrt{\left\langle b_k^2 \right\rangle}$ is the Kuhn segment length, $\beta$ is a numerical coefficient of order one, depending on the details of the model. The relation implies strong coupling between rotation and spatial displacement at the time interval indicated. Therefore, as can be seen from expressions (11), (16) and (17), the intramolecular relaxation contribution at times $\tau_s \ll t \ll \tau_1$ is decaying as the fourth power of thermal spatial displacements:

$$A_0^{s;isot}(t) = \frac{5}{12\pi^3} \tilde{S} \beta \frac{b^8}{\left\langle r_k^2(t) \right\rangle^2}. \tag{18}$$

Comparing this expression with (10) we see that in isotropic polymer models intrasegmental contribution is decaying with time more rapidly than the intermolecular contribution.

The situation is different for the case of the tube-reptation model. At times $\tau_e \ll t \ll \tau_1$, the number of Kuhn segments between two entanglements $N_e$ is the main phenomenological parameter of the tube-reptation model. It is connected with the tube diameter by the relation $d = N_e^{1/2} b$. In this model, as said, motion of the polymer chain is strongly anisotropic, i.e. strongly correlated with the initial chain conformation, because it is performed inside a tube assuming a Gaussian conformation. In the interval $\tau_e \ll t \ll \tau_1$, all discussed time correlation functions decay proportionally to the probability of a polymer segment to return to the initial part of the tube [7,22]. For the discussed correlation function this has the following consequence:

$$\left\langle \left[\left(b_k^2(t)\right) - 3\left(b_k^z(t)\right)^2\right]\left[\left(b_k^2(0)\right) - 3\left(b_k^z(0)\right)^2\right]\right\rangle^{rep} \propto \frac{4}{3}\left\langle \vec{b}_k(\tau_e)\vec{b}_k(0)\right\rangle^2 P(t), \quad (19)$$

where $P(t)$ is the probability for the polymer chain to return to the initial part of the tube and, of course $t \gg \tau_e$, and the pre-factor $\frac{4}{3}\left\langle \vec{b}_k(\tau_e)\vec{b}_k(0)\right\rangle^2$ takes into account that for times $t \leq \tau_e$, motion of the chain is Rouse-like, i.e. isotropic. The probability $P(t)$ in the tube-reptation model is inversely proportional to the segmental mean squared displacement:

$$P(t) \propto \frac{d^2}{\left\langle r_k^2(t)\right\rangle}. \quad (20)$$

Now it is possible to see that in the tube-reptation model the intrasegmental contribution decays inversely to the segmental mean square displacement:

$$A_0^{s;rep}(t) \propto \frac{5}{12\pi^3}\tilde{S}\beta \frac{b^8}{d^2\left\langle r_k^2(t)\right\rangle}. \quad (21)$$

In conclusion the different polymer models (isotropic vs anisotropic) differ in the way they relate $A_0(t)$ with the segmental mean square displacement eqs. (18 vs 21). Finally, we note that the segment-segment intramolecular contribution can be neglected in most realistic situations [21].

3. **New possibilities for proton NMR experiments studying polymer dynamics in melts**

The Inverse Fourier transforms of the general expression for the proton spin-lattice relaxation rate connects the total dipolar correlation function with the dispersion of the spin-lattice relaxation rate $\frac{1}{T_1(\omega)}$ in the following way:

$$A_1(t) + 2A_2(t/2) = \left(\frac{4\pi}{\mu_0}\right)^2 \frac{5}{3\pi^2} \frac{1}{\gamma_H^4 \hbar^2} \int_0^\infty \frac{\cos(\omega t)}{T_1(\omega)} d\omega. \quad (22)$$

This constitutes the sum of inter- and intramolecular components.

Equation (10) allows one to establish a connection between the relative mean-squared displacement between two polymer segments from different polymer chains and the intermolecular contribution to the total spin-lattice relaxation rate:

$$\frac{1}{\left\langle \tilde{r}^2(t)\right\rangle^{3/2}}+\frac{2}{\left\langle \tilde{r}^2(t/2)\right\rangle^{3/2}}=\left(\frac{4\pi}{\mu_0}\right)^2 \frac{5}{4}\sqrt{\frac{8}{3\pi^3}}\frac{1}{\gamma_H^4 \hbar^2 n_s}\int_0^{\infty}\frac{\cos(\omega t)}{T_1^{\text{inter}}(\omega)}d\omega. \qquad (23)$$

In polymer melts segmental diffusion is sub-diffusive at times $\tau_s \leq t \leq \tau_1$, and the relative mean-squared displacement scales

$$\left\langle \tilde{r}^2(t)\right\rangle \propto t^{\alpha}, \qquad (24)$$

where the exponent $1/4 \leq \alpha \leq 1/2$ for different polymer models. The expression (23) then can be rewritten as:

$$\frac{1+2^{1+3\alpha/2}}{\left\langle \tilde{r}^2(t)\right\rangle^{3/2}}=\left(\frac{4\pi}{\mu_0}\right)^2 \frac{5}{4}\sqrt{\frac{8}{3\pi^3}}\frac{1}{\gamma_H^4 \hbar^2 n_s}\int_0^{\infty}\frac{\cos(\omega t)}{T_1^{\text{inter}}(\omega)}d\omega. \qquad (25)$$

Note, that the $\alpha$-dependent numerical factor in (25) changes rather slowly for intervals where $\alpha = 1/4 \div 1/2$, i.e. the factor $1+2^{1+3\alpha/2} \simeq 4.4 - 3.6$. For the cases where $\alpha < 2/3$ it is also possible to obtain the following analytical expression [16]:

$$\left\langle \tilde{r}^2\left(t=\frac{1}{\omega}\right)\right\rangle = \left(\left(\frac{\mu_0}{4\pi}\right)^2 \frac{12\sqrt{\pi}}{5\sqrt{6}} f_1(\alpha) \frac{\gamma_H^4 n_s}{\omega} T_1^{\text{inter}}(\omega)\right)^{-2/3}, \qquad (26)$$

where $f_1(\alpha) = \dfrac{\pi\left(1+2\cdot 2^{3\alpha/2}\right)}{2\cos\left(\dfrac{3\pi\alpha}{4}\right)\cos\left(\dfrac{3\alpha}{2}\right)}$

An interesting behavior of the dispersion of the spin-lattice relaxation rate occurs in the limit of small frequencies $\omega \tau_1 \leq 1$. At corresponding times $t \geq 1/\omega$ the intramolecular contribution to the total dipolar correlation function is decaying exponentially:

$$A_p^{\text{intra;rep}} \propto \exp\left\{-\frac{t}{\tau_1}\right\} \qquad (27)$$

for the tube-reptation model and as

$$A_p^{\text{intra;isot}} \propto \exp\left\{-\frac{2t}{\tau_1}\right\} \qquad (28)$$

for the isotropic models. In the same limit, the intermolecular contribution in the same limit is decaying slower for all polymer dynamic models:

$$A_p^{\text{inter}} \propto \frac{1}{\left(4R_g^2 + 12Dt\right)^{3/2}}, \tag{29}$$

where $R_g = \frac{bN^{1/2}}{\sqrt{6}}$ is the macromolecule's radius of gyration and $D$ its self-diffusion coefficient. Therefore in the limit $\omega \tau_1 \leq 1$, the frequency dispersion of the total spin-lattice relaxation rate will always be determined by the intermolecular interactions, a fact well documented by recent experiments with polymer melts [12,13,27-28].

In order to obtain explicit expressions for $\frac{1}{T_1(\omega)}$ at low frequencies the intermolecular contribution can be rewritten:

$$\frac{1}{T_1^{\text{inter}}(\omega)} = \left(\frac{\mu_0}{4\pi}\right)^2 \frac{4}{5}\sqrt{\frac{3\pi}{2}} \gamma_H^4 \hbar^2 n_s \int_0^\infty dt \frac{\left(\cos(\omega t) + 4\cos(2\omega t)\right)}{\left\langle \tilde{r}^2(t) \right\rangle^{3/2}} =$$
$$\left(\frac{\mu_0}{4\pi}\right)^2 \frac{4}{5}\sqrt{\frac{3\pi}{2}} \gamma_H^4 \hbar^2 n_s \int_0^\infty dt \left(\frac{1}{\left\langle \tilde{r}^2(t) \right\rangle^{3/2}} + \frac{2}{\left\langle \tilde{r}^2(t/2) \right\rangle^{3/2}}\right)\left(1 - 2\sin^2(\omega t/2)\right). \tag{30}$$

At times about and longer than the terminal relaxation time $t \geq \tau_1$ it is reasonable to expect that motions of two chains separated by distances of the order $R_g$ are nearly independent of each other. Therefore, the relative segmental mean-squared displacements from different macromolecules can be estimated as twice larger than their mean squared displacements:

$$\left\langle \tilde{r}^2(t) \right\rangle = 2\left\langle r_k^2(t) \right\rangle . \tag{31}$$

The segmental mean-squared displacement can be represented by a contribution from thermal motion from different Rouse normal modes:

$$\left\langle r_k^2(t) \right\rangle = 6Dt + \frac{12R_g^2}{\pi^2} \sum_P \frac{1 - \exp(-t/\tau_p)}{p^2}, \tag{32}$$

where $\tau_p$ is the relaxation time of the Rouse normal mode with number $p = 1,...,N-1$. Note that (32) is only formally of the same structure as the Rouse model. It is actually general in nature and can be used for alternative polymer dynamical models as well. Details of models more complicated than the Rouse model, taking into account entanglements effects, are hidden in the molecular mass dependence of the self-diffusion coefficient and the mode-dependence of the relaxation time $\tau_p$.

For example in the Rouse model $D^R = \dfrac{D_0}{N^2}$, $D_0$ is the segmental diffusion coefficient, $\tau_p^{Rouse} = \tau_s \dfrac{N^2}{p^2}$, whereas in the reptation model $\tau_p^{rep} = \tau_s \dfrac{3N^3}{N_e p^2}$, $D^{rep} = \dfrac{N_e}{3N^2} D_0$. For the n-Renormalized Rouse Model in the discussed time interval the Markovian approximation is correct and

$$\tau_p^{nR} = \tau_p^{Rouse}\left\{1+\left(\dfrac{N}{N_e p}\right)^{n/2}\right\}, \quad D^{nR} = D_R\left\{1+\left(\dfrac{N}{N_e}\right)^{n/2}\right\}^{-1}.$$

In the limit $t \gg \tau_1$, expression (32) for all models approaches the following:

$$\langle r_k^2(t)\rangle \simeq 6Dt + 2R_g^2. \tag{33}$$

The second line of expression (30) shows that in the limit $\omega\tau_1 \to 0$, the intermolecular part of the spin-lattice relaxation rate is approaching the frequency independent plateau value:

$$\dfrac{1}{T_1^{inter}(0)} = \left(\dfrac{\mu_0}{4\pi}\right)^2 \dfrac{4}{5}\sqrt{\dfrac{3\pi}{2}}\gamma_H^4 \hbar^2 n_s \int_0^\infty dt \left(\dfrac{1}{\langle \tilde{r}^2(t)\rangle^{3/2}} + \dfrac{2}{\langle \tilde{r}^2(t/2)\rangle^{3/2}}\right). \tag{34}$$

Employing the right hand side of (33) for the approximation (32) we obtain its lower boundary:

$$\dfrac{1}{T_1^{inter}(0)} \geq \left(\dfrac{\mu_0}{4\pi}\right)^2 \sqrt{\dfrac{\pi}{6}} \dfrac{\gamma_H^4 \hbar^2 n_s}{R_g D}. \tag{35}$$

The first correction to the plateau value at the limit $\omega\tau_1 \to 0$ can be obtained from (30), when one approximates relative mean squared displacements $\langle \tilde{r}^2(t)\rangle$ in its frequency dependent part, containing $\sin^2(\omega t/2)$, by expression (33) in which the term $2R_g^2$ is neglected. Then integration can be performed exactly:

$$\dfrac{1}{T_1^{inter}(0)} - \dfrac{1}{T_1^{inter}(\omega)} = \left(\dfrac{\mu_0}{4\pi}\right)^2 \dfrac{\pi}{30}\left(1+4\sqrt{2}\right)\gamma_H^4 \hbar^2 n_s \dfrac{\sqrt{\omega}}{D^{3/2}}. \tag{36}$$

Note that expression (36) has the same formal structure as its analogue obtained many decades ago for the case of low molecular liquids using a different formalism [46-49**Fehler! Verweisquelle konnte nicht gefunden werden.**] and which has recently be applied for extracting the diffusion coefficient $D$ using the new possibilities of FFC $^1$H NMR in simple liquids [13-48-52] as well as in polymers [24-25]. With the knowledge of $D(M)$ obtained by applying (36) one may extract $R_g(M)$ from (35), a very interesting possibility.

## 4. Discussion.

In polymer melts at very short times $t \ll \tau_s$ the main contribution to the total correlation function $A_p(t)$ stems from intramolecular protons spin mainly belonging to the same Kuhn segment. At longer times, however, the situation changes significantly. In polymer physics it is established [42-44] that at times $\tau_s \leq t \leq \tau_e$ the polymer dynamics is well described by the Rouse dynamics, which is essentially isotropic. Empirically it is also well known that $N_e \approx 20 \div 40$ for flexible polymer melts [41-44,54]. In this time interval – as follows from expressions (10) and (18) – the intermolecular part $A_p^{inter}(t)$ is decaying slower with time than the intramolecular part $A_p^{intra}(t)$. Apart from that, at times similar to the segmental relaxation time $\tau_s$, the intramolecular part is already reduced by the fast local intra-segmental conformation fluctuations, which is reflected in expressions (13) and (18) through the spin-segment coupling parameter $S_{ij} < 1$. In this time interval (Rouse dynamics) the ratio of the inter- and intramolecular contribution will grow with time as:

$$\frac{A_p^{inter}(t)}{A_p^{intra}(t)} \approx \frac{2\pi^2}{5}\sqrt{3\pi}\left(\frac{r_{ij}}{b}\right)^6 \frac{n_s b^2 \langle r_k^2(t) \rangle^{1/2}}{S_{ij}}. \tag{37}$$

The distance between two nearest spins from the same Kuhn segment typically has values about $r_{ij} \approx 1.3 \div 1.5 \overset{0}{A}$, while the Kuhn segment length in flexible polymers is $b \cong 5 - 20$ A. Therefore, the quantity $\left(\frac{r_{ij}}{b}\right)^6 \ll 1$ is always extremely small. At times on the order of the entanglement time $\tau_e$ the segmental mean-squared displacement becomes about the square of the characteristic length connected with the entanglement effects, which in the tube reptation terminology is known as the tube diameter $\langle r_k^2(\tau_e) \rangle^{1/2} \approx d \approx 30 \div 50 \overset{0}{A}$. This quantity is sufficiently large to render the ratio $\frac{A_p^{inter}(t)}{A_p^{intra}(t)}$ about unity and even larger.

At times $\tau_e \ll t \ll \tau_1$, (entanglement regime) the behavior of the ratio $\dfrac{A_p^{\mathrm{inter}}(t)}{A_p^{\mathrm{intra}}(t)}$ for the tube-reptation model and the isotropic polymer dynamics is different. For isotropic models, (37) is still correct and entanglement effects would be reflected only through changing the time dependence of $\langle r_k^2(t) \rangle^{1/2}$. In the tube-reptation model it decrease in the following way:

$$\frac{A_p^{\mathrm{inter;rep}}(t)}{A_p^{\mathrm{intra;rep}}(t)} \approx \frac{2\pi^2}{5}\sqrt{3\pi}\left(\frac{r_{ij}}{b}\right)^6 \frac{n_s b^2 d^2}{S_{ij}\langle r_k^2(t)\rangle^{1/2}}. \tag{38}$$

For times $t \gg \tau_1$ the ratio will increase with time for all models, although slightly differently. For isotropic models it follows:

$$\frac{A_p^{\mathrm{inter;isot}}(t)}{A_p^{\mathrm{intra;isot}}(t)} \approx \frac{2\pi^2}{5}\sqrt{3\pi}\left(\frac{r_{ij}}{b}\right)^6 \frac{n_s b^2 R_g^4}{S_{ij}\langle r_k^2(t)\rangle^{3/2}}\exp\left\{\frac{2t}{\tau_1}\right\}, \tag{39}$$

while for the tube-reptation model it can be approximated by:

$$\frac{A_p^{\mathrm{inter;rep}}(t)}{A_p^{\mathrm{intra;rep}}(t)} \approx \frac{2\pi^2}{5}\sqrt{3\pi}\left(\frac{r_{ij}}{b}\right)^6 \frac{n_s b^2 d^2 R_g^2}{S_{ij}\langle r_k^2(t)\rangle^{3/2}}\exp\left\{\frac{t}{\tau_1}\right\}. \tag{40}$$

Note that both quantities $A_p^{\mathrm{inter}}(t)$ and $A_p^{\mathrm{intra}}(t)$ are decreasing functions of time. The intermolecular contribution $A_p^{\mathrm{inter}}(t)$, as was experimentally demonstrated in the works of the Kimmich et.al. [16,17], can be separated from the total dipolar contribution which opens up new directions for applying proton-deuteron NMR for the experimental investigations polymer dynamics. Later based on the temperature-frequency superposition principle this was extended on essentially more time/ frequency window [12,13,15,18,19]. This will be demonstrated in Section 5.1 where the normalized dipolar correlation function

$$C^{DD}(t/\tau_s) = \frac{A_1(t/\tau_s) + 2A_2(t/2\tau_s)}{A_1(0) + 2A_2(0)} \tag{41}$$

is discussed together with its separation in to an intra- $C_{\mathrm{intra}}(t)$ and an intermolecular part $C_{\mathrm{inter}}(t)$.

## 5  Recent experimental examples

### 5.1  Full dipolar correlation function in polymer melts

As a continuation of the pioneering work of Kimmich and collaborators [7] the dynamics of linear polymers such as 1,4-polybutadiene (PB) with different $M$ has been studied by field-cycling (FC) $^1$H NMR relaxometry [13,15,19,24,55-56]. Transforming the measured relaxation rates $R_1=1/T_1$ to the susceptibility representation $\chi''_{DD} = \omega/T_1$ and applying frequency-temperature superposition (FTS) master curves $\chi''_{DD}(\omega\tau_s)$ have been constructed by shifting individual dispersion data measured at a temperature solely along the frequency axis. The master curve covers the full polymer dynamics including local, Rouse and entanglement contributions. The case for a series of PB is shown in Figure 1a where about ten dispersion curves collected typically in the temperature range 200 – 400 K are combined to provide a master curve for a given M. Figure 1a also includes low-frequency relaxation data obtained by compensating Earth and stray fields [24,25,31,32]. Frequencies down to some 100 Hz can be reached; thus almost two orders of magnitude are gained with respect to a commercial FC NMR spectrometer. Scaling by the segmental correlation time $\tau_s$ yields "isofrictional" spectra and provides a common peak at $\omega\tau_s \approx 1$ representing the primary ($\alpha$-) relaxation (also denoted local or segmental relaxation) governed by the glass transition phenomenon. With increasing $M$ a continuously rising excess intensity on the low-frequency side of the peak ($\omega\tau_s < 1$) is discernible which is due to the slower $M$-dependent polymer dynamics. For the high-$M$ ($M > M_e$) curves three relaxation regimes (0, I, II) are distinguished, and they can be attributed to local (0), Rouse (I) and entanglement dynamics (II).

Fourier-transforming the master curve of the spectral density obtained from susceptibility $\chi''_{DD}(\omega\tau_s)$ allows displaying them as the full dipolar correlation function $C^{DD}(t/\tau_s)$ in Figure 1b. Applying FTS the correlation loss is probed over nine decades in time and eight in amplitude. While the low-$M$ system (PB 466, dotted line) exhibits essentially a stretched exponential decay typical for simple liquids, for higher $M$ the relaxation becomes increasingly retarded. Depending on M characteristic power-laws $t^{-\alpha}$ can be identified (regime I and II). In the time range up to $t/\tau_s < 10^3$ a common envelope with $\alpha = 0.85$ is found which is not altered at high $M$. This is close to $\alpha = 1$ predicted by the Rouse theory [7] expected at short times. Above $M_e$ entanglement dynamics set in leading to correlation loss decaying even slower at longest times at $t > \tau_e$ (regime II). Here, a $M$-dependent power-law $t^{-\alpha(M)}$ is recognized, i.e., the exponent $\alpha$ is reduced with growing $M$ finally reaching $\alpha = 0.32$ which is actually rather close to the prediction of the tube reptation model ($\alpha = 0.25$). For M $\leq$ 56500 the curves at longest times bend down due to terminal relaxation. Similar results have been reported for other polymers like polyisoprene, polypropylene glycol and polydimethylsiloxane [57]. Moreover, the agreement with the results from double quantum (DQ) $^1$H NMR is almost perfect [11]. In the latter, as discussed in the Introduction, similar dipolar fluctuations are probed yet monitored in the time domain. In conclusion, extensive FC as well as DQ $^1$H NMR studies on PB appear to support the tube-reptation.

When comparing the observed power-laws and the corresponding exponents in Fig. 1(b) probed by FC $^1$H NMR to the ones predicted by polymer theories one has to keep in mind that the dipolar correlation function $C^{DD}(t)$ contains both intramolecular (reflecting reorientational dynamics) and intermolecular contributions (related from translational dynamics) discussed in 3.2). Indeed, as anticipated above, it has been shown that the intermolecular contribution to $R_1(\omega)$ must not be ignored suggesting that the exponent $\alpha$ might also be influenced by the intermolecular relaxation [13,16,17,25]. Moreover, the intermolecular relaxation is an important source of information regarding translational dynamics in polymer melts.

In order to separate intra- and intermolecular relaxation contributions mixtures of protonated and deuterated PB and PDMS have been studied, i.e., the isotope dilution technique is applied [25]. In Figure 2 the correlation functions $C_{DD}(t/\tau_s)$, $C_{intra}(t/\tau_s)$, $C_{inter}(t/\tau_s)$, and $C_Q(t)$ (the latter from FC $^2$H NMR) and are displayed for PB with high $M$. At short times where glassy (or "local") dynamics dominate (regime 0) all correlation functions almost coincide. In the Rouse regime (I) weak differences among the correlation functions are observed which become, however, significant in the entanglement regime (II). Whereas the power-law exponent $\alpha$ of $C_{inter}(t/\tau_s)$ is always lower than the corresponding one of $C_{DD}(t/\tau_s)$ that of $C_{intra}(t/\tau_s)$ is always higher. Moreover, the correlation function $C_{intra}(t/\tau_s)$ obtained from the intramolecular contribution agrees well with $C_Q(t)$ from FC $^2$H NMR, the latter by its very nature probing solely reorientational dynamics. Due to the strong intermolecular contribution at low frequencies, the intramolecular part significantly changes with respect to the total relaxation, and it is found that the exponent of $C_{intra}(t/\tau_s)$ in regime II is rather high and does actually not agree with the prediction $\alpha = 0.25$ of the tube-reptation model. Explicitly, while the long-time exponent in the total dipolar correlation $C_{DD}(t)$ is $\alpha = 0.32\pm0.02$ it becomes $\alpha = 0.41\pm0.05$ for $C_{intra}(t/\tau_s)$. We note that the exponent has been slightly corrected with respect to the originally published one [25] applying now a derivative method to determine the smallest exponent in regime II [30]. It appears that at longest times already the terminal relaxation leads to a somewhat steeper decay.

Since the ratio $\dfrac{A_p^{\text{inter}}(t)}{A_p^{\text{intra}}(t)}$ in accordance with eq. (37) increases with decreasing frequency the observed trend may favor an isotropic model of segmental motion. Yet, it is too early to derive a final conclusion based on the available data, and it would be a realistic approach to assume an intermediate scenario in between the existing isotropic and anisotropic models. In particular, the assumption of a universal behavior for all types of linear polymers will be subject to detailed investigation in the near future. Very recently, a DQ $^1$H NMR study [30] has appeared which claims that the exponent $\alpha$ in regime does not change when isotope dilution is applied providing also agreement with simulation data [11,30,59]. This implies that ratio the of intra- and intermolecular correlation is independent of frequency or time (actually a ratio about one has been reported). This large yet time independent ratio is at variance with all models presented so far. In conclusion, concerning the verification of the relation of translational and reorientational dynamics in polymer melts the experimental situation is still not fully settled.

Computer simulations can be very useful for detailed investigations on well characterized model polymer systems. Figure 3 shows a result from atomistic molecular dynamics (MD) simulations of poly(propylene oxide) compared to FC $^1$H NMR results on poly(propylene glycol) [60]. Almost quantitative agreement is found. In the frequency range covered glassy, Rouse and the onset of entanglement dynamics is probed. Here, the influence of the time dependent intermolecular contributions is still small and $C_{\text{DD}}(t/\tau_s) \cong C_{\text{intra}}(t/\tau_s)$ still holds. Even separating the intra- and intermolecular contributions to the total dipolar correlation function becomes now available [61]. It is clearly shown that the intermolecular contribution $A_p^{\text{inter}}(t)$ decays more slowly with increasing time compared to the intramolecular contribution $A_p^{\text{inter}}(t)$, in particular, at longest times the power-law $C_{\text{DD}}(t/\tau_s) \propto t^{-3/2}$ characteristic of free diffusion is well documented. It would be very important to extend the investigated molecular mass range, even though this is complicated by the fact that the intermolecular contribution is of long-range nature and therefore demands to take into account interactions of a large number of spins. This situation can possibly be simplified in part by using a coarse-grained polymer melts model.

## 5.2 Extracting the segmental means square displacement from intermolecular relaxation

Using relations (10), (23) and (26) it is possible to extract the relative mean squared displacement from experimental relaxation data from $\frac{1}{T_1^{\text{inter}}(\omega)}$. Additional proton NMR methods like FID, Hahn Echo, DQ NMR will provide complimentary information. First experimental investigations in this direction have already been published in papers [13,16,17,25]. The result of PB for two $M$ values is shown in Figure 4. Clearly two regimes are recognized. The first one, at short times, yields a power-law $t^{0.49\pm0.03}$ for both $M$ in accordance with the Rouse model prediction of $t^{0.5}$. At long times a power-law is observed for the high-$M$ PB which is close to $t^{0.25}$ expected for the constrained Rouse dynamics (regime II), whilst PB 24300 shows a tendency to crossover to a similar behavior but free diffusion interferes at the longest time. Comparable results are found for PDMS [25]. Thus, the mean squared displacement can equally well be obtained from FC NMR as in the case of neutron scattering [62].

## 5.3 Diffusion coefficient extracted from the low-frequency dispersion of the total proton relaxation

The use of expression (36) appears very promising since it allows, without the need of applying the isotope dilution technique, to measure the time-dependence of the self-diffusion coefficient of polymer chains and of its molecular mass dependence in the limit $\omega\tau_1 \to 0$, i.e. $D(T,M)$. First measurements of this kind have been published in [27,28]. This application of proton NMR relaxometry can potentially circumvent some limitations of pulsed field gradient (PFG) or static field gradient NMR for measurements of the self-diffusion coefficient, which is ultimately limited by the spin-diffusion process [58]. In addition, the analysis of the dispersion data is straightforward and allows one to collect a wealth of data within short measuring times.

As recently demonstrated for simple (non-polymeric) liquids, the slow translational dynamics probed by the intermolecular relaxation dominates the $^1$H relaxation dispersion at low frequencies [48-52], and this is also the case for polymers as discussed above. Applying eq. (36) thus provides D(T,M) also for polymers [27,28]. Figure 5 shows the results for $D(T,M)$ for the polymers PDMS, PB, and PS. The temperature as well as $M$-dependence can be monitored over a large range. Strong changes with M are observed when the low-M limit of the monomeric liquid is approached, in particular for PS. Good agreement is found with data from field gradient NMR. Actually, Thus, FC $^1$H NMR as an alternative, simple method of determining diffusion coefficients at least in neat systems is established. Extension to binary systems like ionic liquids, are promising [**Fehler! Verweisquelle konnte nicht gefunden werden.**].

In order to get the diffusion coefficient, equation (36) exploits the expansion of the intermolecular relaxation to lowest order in frequency. If one uses second line of (30) and relations (31) and (32), the analysis can be extended up to frequencies $\omega\tau_1 \sim 1$, which allows one to experimentally investigate the details of the transition from anomalous diffusion to normal diffusion.

## 6    Conclusion

We like to state that the systematic investigations of the inter- and intramolecular contribution to proton spin dynamics in polymer melts as well as in other soft matter systems is a new and very promising direction for future investigations. Rotational as well as translational dynamics in terms of segmental reorientation and segmental mean-squared displacement, respectively, are accessible. The frequency/time dependence of the ratio of inter- and intramolecular relaxation allows to discriminate different microscopic dynamics. Employing the field-cycling technique to measure the dispersion of the spin-lattice relaxation together with the application of frequency-temperature superposition and/or in combination with other NMR techniques like FID, Hahn echo or double quantum NMR, an extremely broad dynamic range may be covered. Thereby, NMR will establish itself as a method of molecular rheology.

**Acknowledgments:**

Financial support from Deutsche Forschungsgemeinschaft (DFG) through grants STA 511/13-1 and RO 907/15 and 907/16 is gratefully acknowledged.

**Figure captions**

**Fig. 1:**

(a) Susceptibility master curves as a function of reduced frequency $\omega\tau_s$ for polybutadiene of different molecular weight $M$ as obtained from FC $^1$H NMR. Frequency below which Earth field compensation is applied marked by vertical (red) dashed line. Vertical dotted lines: relaxation regimes 0, I, and II, i.e., glassy dynamics, Rouse and entanglement dynamics, respectively. (b) Dipolar correlation function $C^{DD}(t/\tau_s)$ obtained from the data in (a) compared to those from double quantum (DQ) $^1$H NMR[11]. Dotted curve: low-$M$ system representing glassy dynamics. Dashed line: power-law in Rouse regime (I). Solid line: power-law by the tube-reptation model (adapted from [24]).

**Fig. 2:** Different correlation functions for PB with $M = 24300$ (a) and 196000 (b): $C_{DD}(t/\tau_s)$ (comprising intra- and intermolecular contributions), $C_{intra}(t/\tau_s)$, $C_Q(t/\tau_s)$ (from FC $^2$H NMR), and $C_{inter}(t/\tau_s)$. For regime II their power-law exponents $\alpha$ are indicated (adapted from [25]).

**Fig. 3:** Comparison of the correlation function $C_{DD}(t/\tau_s)$ from MD simulations of poly(propylene oxide) and $C_{intra}(t/\tau_s)=C_2((t/\tau_s))$ FC $^1$H NMR results of poly(propylene glycol). Three different dynamic regimes (0, I, II) can be identified, namely glassy, Rouse and the onset of entanglement dynamics, respectively (with permission from [60]).

**Fig.4:** Segmental mean squared displacement $<\mathbf{R}^2(t/\tau_s)>$ for polybutadiene with $M = 24300$ and 196000 calculated from the frequency dependence of the spin-lattice relaxation rate according to eq. (26) (adapted from ref. [25])

**Fig.5:** Diffusion coefficients $D(T,M)$ obtained from FC $^1$H NMR as a function of inverse temperature for (a) poly(dimethylsiloxane) (PDMS); (b) 1,4-poly(butadiene) (PB), and (c)

poly(styrene) (PS) with molecular masses *M* in g/mol as indicated. Solid lines: Vogel-Fulcher-Tammann interpolation (cf. [28]).

Fig. 1.

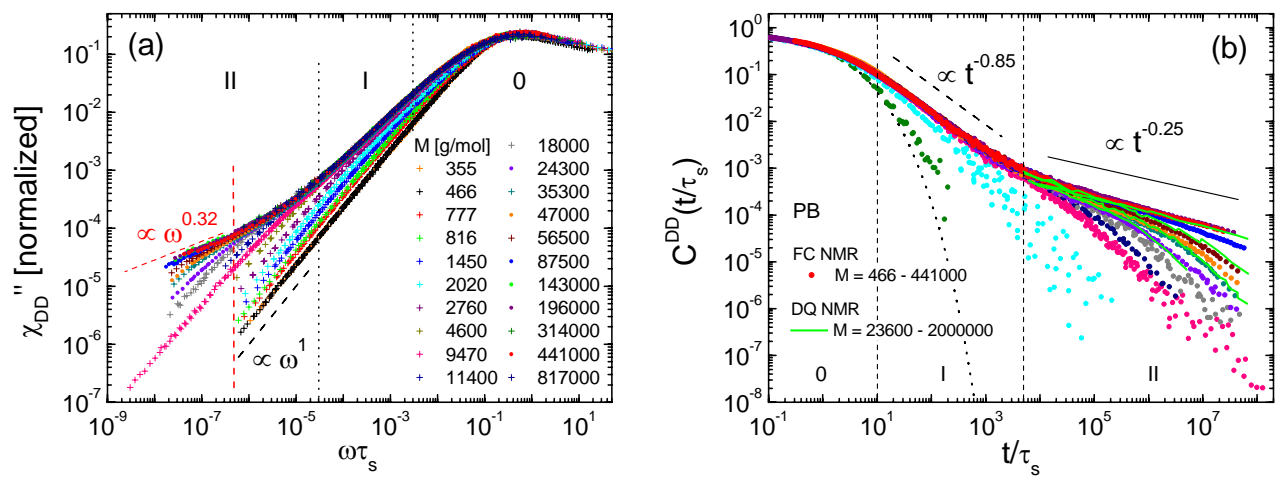

Fig. 2.

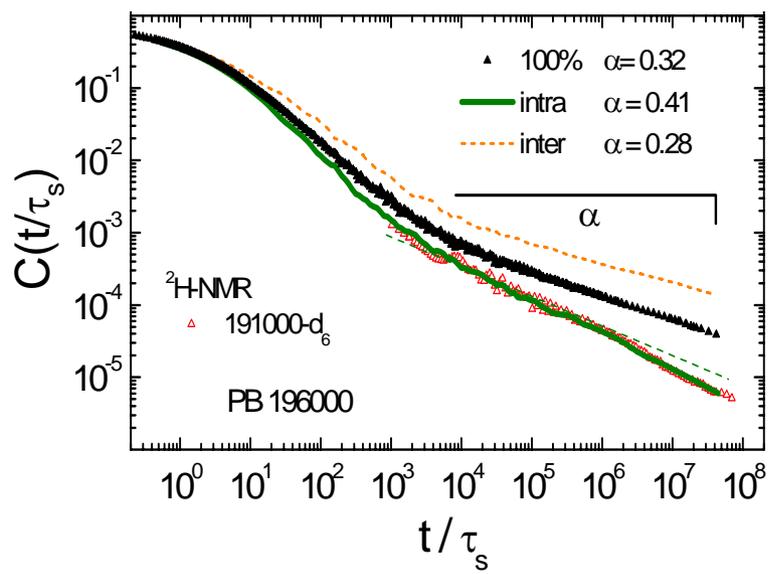

Fig. 3.

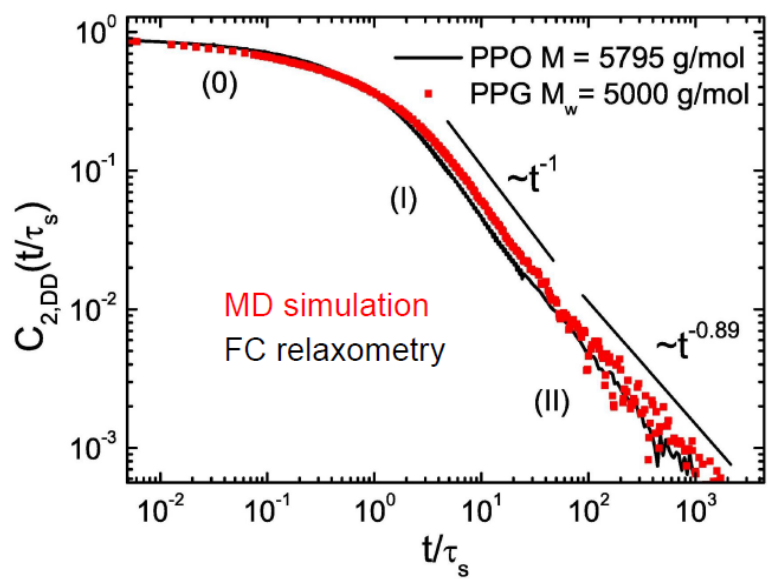

Fig. 4.

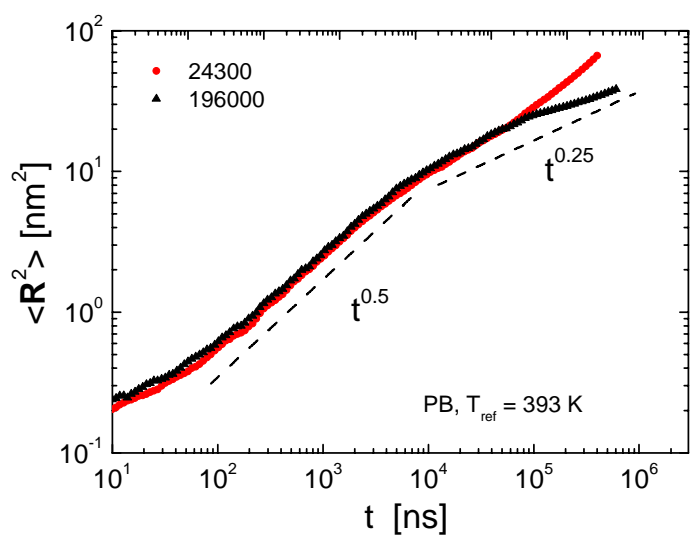

Fig. 5.

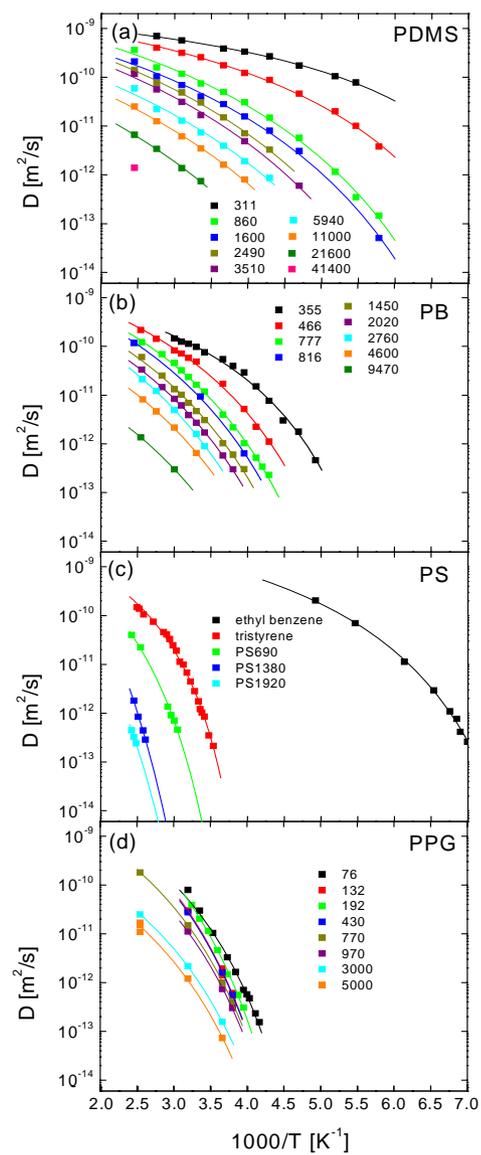

28